\documentclass[sn-mathphys, twocolumn]{sn-jnl}

\jyear{2022}%

\theoremstyle{thmstyleone}%

\theoremstyle{thmstyletwo}%

\theoremstyle{thmstylethree}%

\renewcommand{\footnoterule}{%
  \kern -3pt
  \hrule width .49\textwidth height 1pt
  \kern 2pt
}

\begin{document}

\title[Exo-SIR]{Exo-SIR: An Epidemiological Model to Analyze the Impact of Exogenous Spread of Infection}

\author*[1]{\fnm{Nirmal Kumar} \sur{Sivaraman}}\email{nirmal.sivaraman@lnmiit.ac.in}

\author[2]{\fnm{Manas} \sur{Gaur}}\email{MGAUR@email.sc.edu}

\author[1]{\fnm{Shivansh} \sur{Baijal}}\email{shivanshbaijal@gmail.com}

\author[1]{\fnm{Sakthi Balan} \sur{Muthiah}}\email{sakthi.balan@lnmiit.ac.in}

\author[2]{\fnm{Amit} \sur{Sheth}}\email{AMIT@sc.edu}

\affil[1]{\orgdiv{Department of Computer Science and Engineering}, \orgname{The LNM Institute of Information Technology}, \orgaddress{ \city{Jaipur}, \country{India}}}

\affil[2]{\orgdiv{AI Institute}, \orgname{University of South Carolina}, \orgaddress{\country{USA}}}

\abstract{Epidemics like Covid-19 and Ebola have impacted people's lives significantly. The impact of mobility of people across the countries or states in the spread of epidemics has been significant. The spread of disease due to factors local to the population under consideration is termed the endogenous spread. The spread due to external factors like migration, mobility, etc. is called the exogenous spread. In this paper, we introduce the Exo-SIR model, an extension of the popular SIR model and a few variants of the model. The novelty in our model is that it captures both the exogenous and endogenous spread of the virus. First, we present an analytical study. Second, we simulate the Exo-SIR model with and without assuming contact network for the population. Third, we implement the Exo-SIR model on real datasets regarding Covid-19 and Ebola. We found that endogenous infection is influenced by exogenous infection. Furthermore, we found that the Exo-SIR model predicts the peak time better than the SIR model. Hence, the Exo-SIR model would be helpful for governments to plan policy interventions at the time of a pandemic. 
}

\keywords{COVID-19, Ebola, Epidemic modeling, Compartment model, Exogenous infection, Endogenous infection, SIR, Exo-SIR}



\maketitle

\section{Introduction}
An epidemic is a disease that spreads rapidly to a large number of people in a given population within a short period. Many epidemics occur in the world. Covid-19 and Ebola are recent prominent examples. 

People have tried many methods to study epidemics. The susceptible, infected, and recovered model (SIR model) is considered as one of the seminal models of epidemics \cite{harko2014exact}. A recent work \cite{cao2021covid} gives a comprehensive review of the methods to model and analyze Covid-19. Out of these methods, the models relevant to our model are compartmental models. They are prominent methods used for the analysis and prediction of Covid-19 dynamics \cite{cao2021covid} \cite{kumar2020review} \cite{kotwal2020predictive}. However, these works consider only infection from people to people from within the population and do not consider any external source of infection explicitly.

World Health Organization (WHO) has identified external transmission as one of the three modes of transmission \cite{world2020critical}. According to WHO, the infection within the population is called as \emph{Local transmission and community transmission}, and the infection external to the population is called as \emph{Imported cases}. We call the infection from a source within the population and external to the population as the endogenous and exogenous spread of infection, respectively. Human migration is one of the prime reasons behind the exogenous spread of infection. 

The governments can intervene to curb the spread of the disease by bringing in policies to stop human mobility. However, the implementations of such intervention policies have a lot of challenges. Social disagreement is an example. Social disagreement means people do not abide by the government's directives. The Tablighi Jamaat\footnote{\url{https://en.wikipedia.org/wiki/Tablighi_Jamaat}} religious congregation that happened in India and the human mobility as a result of it is an example of social disagreement. 

As a part of intervention, the Governments can restrict human mobility. However, they cannot completely prevent all such human mobility and migration. For example, in the Indian sub-continent, people migrate to metropolitan cities for work. Due to the risk of COVID-19 exposure in these overpopulated cities, people migrate back to their homes \cite{rajan2020covid}. This is also known as reverse migration\footnote{\url{https://www.epw.in/journal/2020/19/commentary/migration-and-reverse-migration-age-covid-19.html}}.

The government cannot deny one's right to go home. However, the government can allow necessary movement in a controlled manner. For example, when people move from one state to another, the state governments can issue passes for anyone who is allowed to travel to that state similar to what was practiced by the State of Kerala\footnote{\url{https://www.news18.com/news/auto/covid-19-omicron-kerala-travel-guidelines-for-international-and-domestic-passengers-4525862.html}}. They can identify the incoming people and ensure that they correctly follow the procedures advised by the respective governments. 

These movements will increase the exogenous spread of the infections compared to the ideal condition of sealed borders. To find the amount of infection during this movement and when the peak occurs, authorities need an explicit model that can predict infection through exogenous means. Our model extends the SIR model and explicitly takes care of the amount of exogenous infection and endogenous infections. In the case of the spread of epidemics, even if there is a small increase in the number of infected people, the impact grows exponentially with time. Hence, it is important to consider the exogenous infections while studying the dynamics of epidemics. This allows the governments to have pertinent information regarding the possible exogenous infections. This gives the government authorities time to prepare their medical resources accordingly.

There are challenges even if people do not migrate. For example, front-line workers like doctors and nurses are more frequently exposed to the virus than a common man. Correspondingly, we need to be able to model different rates of infections for different groups of people. The governments will have to make all the necessary safety equipment available to the front-line workers and monitor their health constantly to control the infection as a measure of intervention.

In this context, we address the following research questions that significantly modifies the current, well-studied SIR model:

\begin{enumerate}
	\item How to quantify the exogenous spread of infection?
	\item What is the interplay between the exogenous and endogenous spread of infection concerning the following:
	
	\begin{enumerate}
		\item In the presence of social disagreement.
		\item In the presence of controlled migration.
		\item In the presence of \emph{n communities} that have a different rate of infection - e.g., front line workers such as healthcare workers or hospitality workers.
	\end{enumerate}
	
	\item What is the change in the peak position (the most significant number of people infected in a unit of time) in the presence of exogenous infection?
	\item What is the change in the height of the peak in the presence of exogenous infection?
\end{enumerate}

The following are our contributions in this work. We study the impact of external reasons of infections such as cross-border mobility on COVID infection by introducing a novel SIR-like compartmental model called Exo-SIR. 

We study three variants of the model applicable for special scenarios like the presence of social disagreement, the presence of different groups that have a different amount of risk, and infectiousness like the front line workers. 

We analyze the interplay between endogenous and exogenous infections during the Covid-19 and Ebola pandemics in the following ways. 
    \begin{enumerate}
        \item Analytically. 
        \item By simulating the Exo-SIR model with and without assuming contact network for the population. 
        \item By implementing the Exo-SIR model on real datasets regarding Covid-19 and Ebola.
    \end{enumerate}
We compare the predictions of Exo-SIR with the SIR model using real data on the recent spread of the Covid-19 in India and the USA and the spread of Ebola in Africa as the ground truth. 

This paper is structured as follows. Section~\ref{Re}, discuss related works and preliminaries. Then, we formulate the Exo-SIR model by extending the SIR model and discuss the different variants of the model (in Section~\ref{the_model}). We analyse our model by comparing it with the SIR model and study the behavior of the infected population in the presence and absence of exogenous infection (in Section~\ref{analysis}). Then we describe the simulation study where we simulated the SIR model and Exo-SIR model and compared them (in Section~\ref{simulation}). Finally, we study the real data of Covid-19 and Ebola epidemics (in Section~\ref{realData}).

\section{Related works}
\label{Re}
Here, we discuss the works related to the idea of exogenous influence to the population under study.

The work in \cite{zhou2014clinical} considers exogenous infections for Malaria at China - Myanmar border. However, the model is not deterministic. In a deterministic model, individuals in the population are assigned to different subgroups or compartments, each representing a specific epidemic stage. Deterministic models often provide useful ways of gaining sufficient understanding about the dynamics of populations whenever they are large enough \cite{brauer2012mathematical}. Also, the deterministic models are simpler and more popular \cite{tolles2020modeling} \cite{ walker2020impact}. Our model is deterministic.

\subsection{Models of external influence on online social networks}
Information diffusion in online social networks is similar to the way the virus spreads in a population \cite{kumar2021information}. There are a few recent works in the literature that attempt to model the external influence in information diffusion in online social networks \cite{myers2012information}. Moreover, \cite{myers2012information} and \cite{li2015measuring} propose information diffusion model on the network. These works assume that the information flows through an underlining network. Also, they consider links from other websites like the mainstream media as external sources of information. Internal diffusion is when the shared messages do not have any external links. 

The work described in \cite{myers2012information} uses very specific parameters like the following:
\begin{itemize}
	\item probability of any node receiving exposure at time $t$
	\item the random amount of time it takes an infected node to expose its neighbors
	\item how the probability of infection changes with each exposure
	\item the probability that a node $I$ have received $n$ exposures by time $t$
\end{itemize}

The work described in \cite{li2015measuring} traces the information cascade and thereby tries to reconstruct the underlying graph structure as much as possible. Also, they conclude that external influence has a bigger impact on the network when compared to the influence of social media influencers.

\begin{figure}[htbp]
	\centering
	\includegraphics[width=.5\textwidth]{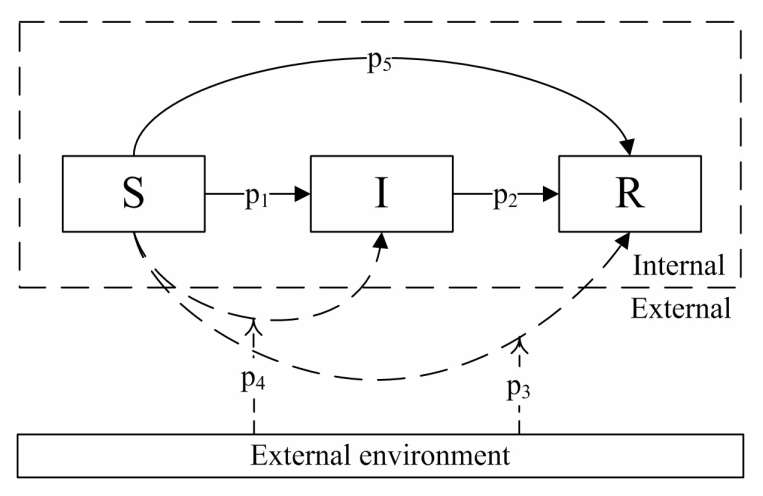}
	\caption{State transition diagram of the diffusion mechanisms in Yang et al's model. Diagram taken from \cite{yang2019modeling}.}
	\label{f1}
\end{figure}

The model that is closest to our work is Yang et. al.'s model \cite{yang2019modeling}. This model is an extends the SIR model (explained in Section~\ref{sir}) by including the external influence on the network. State transition diagram of the diffusion mechanisms of this model is given in Figure~\ref{f1}.

This model is defined in the following way.

\begin{equation}
s+i+r = 1
\end{equation}

\begin{equation}
\frac{ds}{dt} = -p_{1}ksi-((1-p_1)p_3+p_4)\theta s
\end{equation}

\begin{equation}
\begin{split}
    \frac{ds}{dt} = -p_{1}ksi-((1-p_1)p_3+p_4)\theta s \\
-(1-p_1)p_5ksi
\end{split}
\end{equation}

\begin{equation}
\frac{di}{dt} = p_{1}ksi+p_4\theta s-p_2i
\end{equation}

\begin{equation}
\frac{dr}{dt} = p_2i+(1-p_4)p_3\theta s+(1-p_1)p_5ksi
\end{equation}

In Figure~\ref{f1} There are two possible transitions from the state S to I. One path is the normal endogenous path, and the second is due to external influence. These transitions have probabilities $p_1$ and $p4$, respectively. Similarly, there are two possible transitions from the state S to R -- one through endogenous and the other through external influence. Their probabilities are $p_5$ and $p_3$, respectively. However, the transition from the state I to R is not affected by external influence(s).

Although the exogenous infection is modeled in Yang et al.'s model, it fails to capture the dynamics between endogenous and exogenous infections. This is because they do not differentiate between the infections due to exogenous factors from those due to endogenous factors.

\subsection{Other studies of endogenous and exogenous information diffusion}

The dual nature of message flow over the online social network is studied and verified in \cite{de2018demarcating}. Here, the dual nature refers to the injection of exogenous opinions to the network and the endogenous influence-based dynamics. In \cite{fujita2018identifying}, the authors propose a method for extracting the relative contributions of exogenous and endogenous contents. In \cite{agrawal2012learning}, the authors postulate that the nature of the information plays a crucial role in the way it spreads through the network. They quantify two properties of the information -- endogeneity and exogeneity. Endogeneity refers to its tendency to spread primarily through the connections between nodes, and exogeneity refers to its tendency to spread to the nodes, independently of the underlying network. In \cite{oka2014self}, the authors study the bursts that originate from endogenous and exogenous sources and their temporal relationship with baseline fluctuations in the volume of tweets. The study reported in \cite{crane2008robust} classifies the bursts into endogenous and exogenous. According to this study, those bursts that reach the peak almost instantaneously after the diffusion starts and then go down slowly are exogenous bursts. Also, those bursts that gradually increase and slowly decrease are endogenous. 

\subsection{Compartmental models for Covid-19 modeling}
Compartmental models are prominent methods that are used for the analysis and prediction of Covid-19 dynamics. The SIR model is one of the seminal compartmental models. Many compartmental models have come up recently to improve the SIR model. QSIR model \cite{PERI2021100934} \cite{dandekar2020machine} is an example in which they add an extra state to the standard SIR model that represents the number of people in Quarantine. SPCIRD model \cite{zakary2020mathematical} adds three extra states – P, C, and D, where P represents the number of susceptible people who are partially controlled. Partially controlled people are those who can be considered as people not conforming to all the restrictions of the Quarantine. C represents the number of susceptible people who are controlled. Controlled people are those who can be considered as people conforming to all the restrictions of the Quarantine. D represents the number of people who died. Multiple epidemic wave model \cite{kaxiras2020multiple} as its name suggests models the multiple waves of infection that could occur. Time-dependent SIR model \cite{chen2020time} \cite{jung2020real} considers the constants in the SIR model - beta and gamma to be varying with time. However, none of these models consider infections arising from outside the population, mostly due to the cross-border mobility of infected people. Hence, we introduce the Exo-SIR model to address this particular issue.

\subsection{SIR Model}
\label{sir}

This section briefly reviews the SIR epidemiological model to learn how epidemics spread through population. SIR is often used to study information diffusion by approximating the process of epidemic spread. 

In this model, the population is classified into three -- Susceptible (who are prone to infection), Infected (who contain the infection), and Recovered (who do not have the infection and its associated symptoms). In the limit of sizeable total population $N$ that does not change over time, the given equations model the dynamics of the spread \cite{bailey1975mathematical}: 

\begin{equation}
s(t)+i(t)+r(t) = 1
\end{equation}

\begin{equation}
\frac{ds}{dt} = -\beta si
\end{equation}

\begin{equation}
\frac{di}{dt} = \beta si - \gamma i
\end{equation}

\begin{equation}
\frac{dr}{dt} = \gamma i
\end{equation}

\noindent where the fraction of Susceptible, Infected and Recovered people at time $t$ are represented by $s(t),  i(t)$ and $r(t)$ respectively. $\beta$ is the rate of infection, and $\gamma$ is the rate of recovery.

\section{The Model}
\label{the_model}
In this section, we propose the Exo-SIR model. It differs from SIR model in the following ways. It classifies infected nodes into two different types -- Infected from exogenous source and Infected from endogenous source. It also differentiates between the spread from endogenous and exogenous sources. 

Susceptible nodes become infected with a certain probability called the rate of infection. This rate could be different for endogenous and exogenous infections. The nodes affected by endogenous and exogenous sources move into different states. We assume that susceptible nodes get infected from only one of these sources and never from both sources. Hence, even when some nodes are susceptible to endogenous and exogenous infection, they become infected by either an endogenous or an exogenous source but not both. The infected nodes recover with a certain probability called the recovery rate. These nodes move into the recovered state. The advantage of the Exo-SIR model compared to the SIR model is that we can observe the endogenous and exogenous diffusion separately.

We use the following notations:
\begin{itemize}
	\item[$S$] state of susceptible
	\item[$I_x$] state of infected from exogenous source
	\item[$I_e$] state of infected from endogenous source
	\item[$R$] state of recovered
	\item[$i_x$] Fraction of nodes that are infected from exogenous source
	\item[$i_e$] Fraction of nodes that are infected from endogenous source
	\item[$r$] Fraction of nodes that are recovered
	\item[$\beta_x$] Rate at which the exogenous source infects the nodes
	\item[$\beta_e$] Rate at which the nodes infects other nodes
	\item[$\gamma$] Rate at which the nodes get recovered
	\item[] We use the words infection, diffusion and spread interchangeably according to the context.
\end{itemize}

\begin{figure}[htbp]
	\centering
	\includegraphics[width=0.5\textwidth]{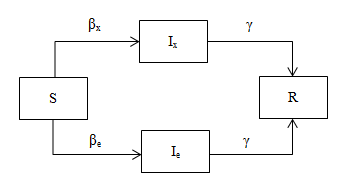}
	\caption{State transition diagram of the nodes in the Exo-SIR model}
	\label{f4}
\end{figure}

The state transition diagram of the Exo-SIR model is given in Figure~\ref{f4}.

We classify infected nodes into two different types -- infected from exogenous source $i_x$ and infected from endogenous source $i_e$.
\begin{equation}\label{key2}
	i_e+i_x = i
\end{equation}

We assume that the total population remains constant.
\begin{equation}\label{key3}
	s+i+r = 1
\end{equation}

A fraction of the susceptible people $s$ gets infected by exogenous sources, and another fraction of $s$ gets infected by endogenous sources. For endogenous infection, the population that is infected plays a big role. Hence, we have
\begin{equation}\label{key4}
	\frac{ds}{dt} =  -\beta_x s -\beta_e s i
\end{equation}

Increase in $i_{x}$ is determined by the number of susceptible nodes and the decrease in $i_x$ is determined by $i_x$. This gives

\begin{equation}\label{key9}
	\frac{di_x}{dt} = \beta_x s - \gamma i_x
\end{equation}

Increase in $i_{e}$ is determined by the number of susceptible nodes and the number of infected nodes and the decrease in $i_e$ is determined by $i_e$. This gives

\begin{equation}\label{key10}
	\frac{di_e}{dt} = \beta_e s i - \gamma i_e
\end{equation}

Increase in $r$ is determined by the number of infected people in the network. This gives

\begin{equation}\label{11}
	\frac{dr}{dt} = \gamma i
\end{equation}

\subsection{Variants of the model to address specific situations}
\label{variants}

In this section, we discuss how the Exo-SIR may be used in the different situations.

\subsubsection{Exo-SIR Model with social disagreement}

This scenario occurs when people do not abide by the government's orders. For example, not wearing masks, not following social distancing, etc. As a result, more people contract the virus, and hence the infectiousness of the disease will go up. This can be represented in the Exo-SIR model by increasing the $\beta_e$ value.

\subsubsection{Exo-SIR Model with people migrating with the permission of the Government}

This scenario can be studied using the Exo-SIR model. Here, we assume that when the government allows people to travel, the government makes sure that these people are isolated and given treatment. Change in $i_{x}$ is influenced by the action of the government that allowed people to travel across their border. Hence, planning and execution efficiency to minimize the impact are essential. This is captured in $\beta_x$. If the government efficiently contains the infection from these people, then the value of $\beta_x$ goes down.

\subsubsection{Exo-SIR model with multiple groups that have different risk of infection}

\begin{figure}[htbp]
	\centering
	\includegraphics[width=0.5\textwidth]{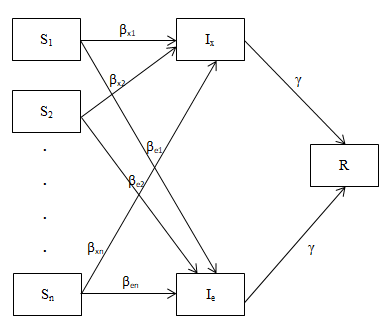}
	\caption{State transition diagram of the model.}
	\label{f4.1}
\end{figure}

This case may be depicted as shown in Figure~\ref{f4.1}. In this case, there are $n$ different groups of susceptible people with varying levels of infection risk. Hence, we add them up wherever we use $s$ in the equations of the Exo-SIR model. Also, the value of each parameter is different for a different group of people. Hence, we have different values for each group of people for the parameters. Hence, there will be the summation of the $n$ groups and parameters for each group. Figure~\ref{f4.1} is the state diagram and the equations are given below. 

\begin{equation}
i_e+i_x = i
\end{equation}

\begin{equation}
s = \sum_{k=1}^{n}s_k
\end{equation}

\begin{equation}
s+i+r = 1
\end{equation}

\begin{equation}
\frac{ds}{dt} =  -\sum_{k=1}^{n}\beta_{xk} s_k - \sum_{k=1}^{n}\beta_{ek} s_k i
\end{equation}

\begin{equation}
\frac{di_x}{dt} = \sum_{k=1}^{n}\beta_{xk} s_k - \gamma i_x
\end{equation}

\begin{equation}
\frac{di_e}{dt} = \sum_{k=1}^{n}\beta_{ek} s_k i - \gamma i_e
\end{equation}

\begin{equation}
\frac{dr}{dt} = \gamma i
\end{equation}

\section{Analysis}
\label{analysis}
In this section, we compare our model with SIR model and analyse the dynamics of exogenous spread and endogenous spread.

\subsection{Comparison with SIR model}
Mirroring the rate of change of $s(t),\ i(t),$ and $r(t)$ in the SIR model (Section~\ref{Re}), we find the expressions for the rate of change of $s(t),\ i(t),$ and $r(t)$ for the Exo-SIR model. 

Rate of change of $s$ is given by

\begin{equation}\label{key8}
	\frac{ds}{dt} = -\beta_x s-\beta_e s i
\end{equation}

Rate of Change of $r$ is given by

\begin{equation}\label{17}
	\frac{dr}{dt} = \gamma i
\end{equation}

Differentiating Eq~\ref{key2} with respect to time, we get

\begin{equation}\label{key11}
	\frac{di}{dt} = \frac{di_e}{dt}+\frac{di_x}{dt}
\end{equation}

\begin{equation}\label{key12}
	\frac{di}{dt} = \beta_e s i - \gamma i_e + \beta_x s - \gamma i_x
\end{equation}

\begin{equation}\label{key14}
	\frac{di}{dt} = \beta_e s i + \beta_x s - \gamma (i_x + i_e)
\end{equation}

Applying Eq~\ref{key2} on Eq~\ref{key14}, we get
\begin{equation}\label{key15}
	\frac{di}{dt} = \beta_e s (i_x + i_e) + \beta_x s - \gamma (i_x + i_e)
\end{equation}

Here, even if we assume that there are no infected people in the beginning -- i.e. $i_e = 0$ and $i_x = 0$, we get the following.

\begin{equation}\label{key16}
	\frac{di}{dt} =  \beta_x s
\end{equation}

This shows that, unlike the SIR model, the Exo-SIR model explains how an infection starts spreading from the state where no one is infected. SIR model assumes that there is an initial outbreak size  $i_0$. This means $i_0$ people are infected in the beginning and $i_0 > 0$ \cite{hethcote2000mathematics}. Our work addresses this limitation of the SIR model. Note that the Exo-SIR model would behave the same way as the SIR model if we assume that $i_x = 0\ and\ \beta_x = 0$.

\subsection{Dynamics of exogenous spread and endogenous spread}
\label{endo-exo}
In this section, we find the relationship between the cumulative exogenous infections ($i_x$) and the daily endogenous infections ($\frac{di_e}{dt}$).

Applying Eq~\ref{key2} on Eq~\ref{key10}, we get

\begin{equation}
\frac{di_e}{dt} = \beta_e s(i_e+i_x) - \gamma i_e
\end{equation}

\begin{equation}
\frac{di_e}{dt}\bigg\rvert_{i_x>0} = \beta_e s(i_e+i_x) - \gamma i_e
\end{equation}

At $i_x = 0$, 
\begin{equation}
\frac{di_e}{dt}\bigg\rvert_{i_x=0} = \beta_e si_e - \gamma i_e
\end{equation}

Since all $\beta_e, s, i_e, and\ \gamma$ are positive, 
\begin{equation}
\frac{di_e}{dt}\bigg\rvert_{i_x=0} <  \frac{di_e}{dt}\bigg\rvert_{i_x>0}
\end{equation}

This shows that $\frac{di_e}{dt}$ increases in the presence of $i_x$. In other words, this shows that the presence of exogenous diffusion causes endogenous diffusion to increase. 

\section{Simulation}
\label{simulation}
We simulate the Exo-SIR model to determine its behavior for various scenarios that are represented by the different values of its parameters. We simulated the model in two ways:

One, by assuming no network (well-mixed population). In this scenario, a susceptible node can get infected from any of the infected nodes in the population under consideration.

Two, By assuming that the people network is a scale-free network. Within this network, the susceptible nodes can catch the infection from only those infected nodes, which they are connected to through an edge, i.e., their immediate neighbors. We chose scale-free network because there are pieces of evidence that the human disease network could be scale-free \cite{szabo2020propagation}. The results of these simulations are discussed in the following section.

\subsection{Using scale-free network}
The analysis presented in this section has been done considering a scale-free contact network for the population under study, which is called Barab{\'a}si-Albert network \cite{barabasi2013network}. Under this scenario, the susceptible nodes can catch the infection from only those infected nodes, which they are connected to through an edge, i.e., their immediate neighbors. We have predicted the values for various combinations of $\beta_x$, $\beta_e$, and $\gamma$ using the Exo-SIR model in the network mentioned above.

Next, we study the dependency of endogenous spread on the exogenous factors through simulation. The step-by-step methodology adopted to carry out the simulation and the analysis is given in Algorithm~\ref{a3}.

\begin{algorithm}[htbp]
	\begin{algorithmic}[1]
		\State Initialize $\beta_x$, $\beta_e$ and $\gamma$ with 3 different values, i.e., 0.1, 0.5 and 0.9. Henceforth, we have 27 different combination of these parameters.
		\State For each of the combinations of $\beta_x$, $\beta_e$, and $\gamma$, iterate over steps 3 and 4 fifty times.
		\State Setup a Barab{\'a}si-Albert network of 1000000 nodes having an average node degree of 2 \cite{barabasi2013network}.
		\State Simulate and predict the values of S, $I_e$, $I_x$, and R using the Exo-SIR model. 
		\State Extract the values of the height of the peak (we call it as peak value) and the time slice at which it occurs (we call it as peak tick), for both endogenous and exogenous peaks from each of the simulations.
		\State Calculate the mean peak value and peak tick of Exogenous and Endogenous infections so that we have one value per combination of $\beta_x$, $\beta_e$, and $\gamma$.
	\end{algorithmic}
	\caption{Algorithm to perform the simulations and analysis by assuming that the contact network in the population is scale-free }\label{a3}
\end{algorithm}

In the above algorithm, we have carried out 50 simulations for each combination of the parameters and averaged it out to address the bias that might get introduced due to the network structure since the setting up of a network in step 3 in the above algorithm is random each time.

\begin{figure}[htbp]
	\centering
	\includegraphics[width=.45\textwidth]{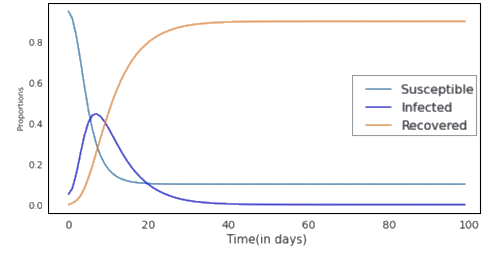}
	\caption{plot of susceptible, infected and recovered with no exogenous source}
	\label{f5}
\end{figure}

\begin{figure}[htbp]
	\centering
	\includegraphics[width=.45\textwidth]{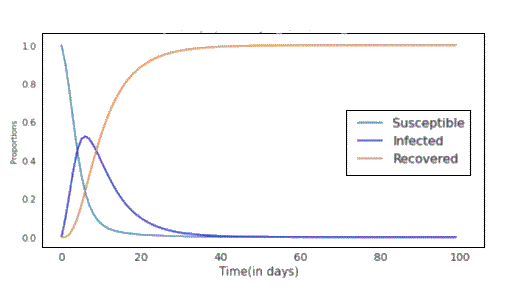}
	\caption{plot of susceptible, infected and recovered with exogenous source}
	\label{f6}
\end{figure}

Sample simulation results are shown in Figures~\ref{f5} and ~\ref{f6}. Figure~\ref{f5} shows the SIR model's simulation results with no exogenous influence, and Figure~\ref{f6} shows the simulation results with exogenous influence. Here we can see that when we consider exogenous factors, the peak of the distribution of the number of the infected population show changes.

\begin{figure}[htbp]
	\centering
	\includegraphics[width=.45\textwidth]{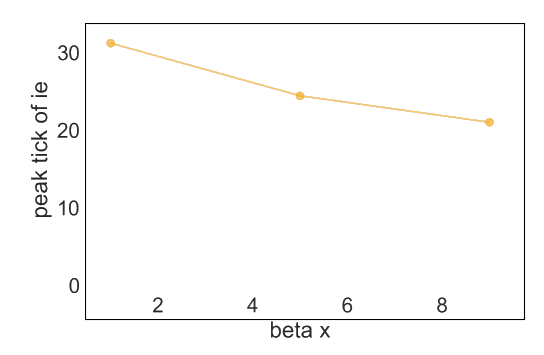}
	\caption{impact of $\beta_x$ on peak tick of $i_e$}
	\label{f200}
\end{figure}

\begin{figure}[htbp]
	\centering
	\includegraphics[width=.45\textwidth]{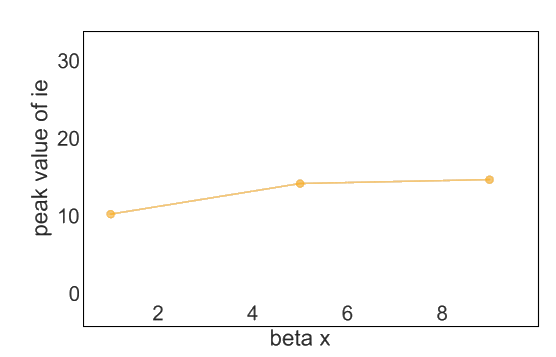}
	\caption{impact of $\beta_x$ on peak value of $i_e$}
	\label{f201}
\end{figure}

Figures~\ref{f200} and ~\ref{f201} are a result of simulation and analysis done as described in Algorithm~\ref{a3} and provide us with the following insights. Figure ~\ref{f200} shows that endogenous peak tick decreases with increase in $\beta_x$. Figure ~\ref{f201} shows that $\beta_x$(exogenous factors) influence the peak value of endogenous infections. The endogenous peak value increases with increase in $\beta_x$.

We can conclude that exogenous source and its infection impacts the endogenous spread in the network by advancing the peak and increasing the height of the peak.


\begin{table}[htbp]
	
	\caption{Impact of $\beta_e$, $\beta_x$ and $\gamma$ on $ln$(\texttt{ie\char`_peak})}
	\label{t0}
	
	\begin{tabular}{|l|l|l|p{.12\textwidth}|}
		\hline
		&&&\\
		& coef & std err & confidence interval \\
		&&&\\
		\hline
		&&&\\
		$\beta_e$ & $0.6319$ & $0.006$ & $0.6204 $ to $0.6434$ \\
		&&&\\
		$\beta_x$ & $0.6319$ & $0.006$ & $0.6204 $ to $0.6434$ \\
		&&&\\
		$\gamma$ & $-0.4390$ & $0.006$ & $-0.4505 $ to $-0.4274$ \\
		&&&\\
		\hline
	\end{tabular}

\end{table}

\subsection{With no network}
In this section, we determine the relative effects of $\beta_x$, $\beta_e$, and $\gamma$ on the endogenous peak statistically and measure the impact of $\beta_x$ on endogenous infections, which is consistent with the results shown above. Here, we did not assume any network for our population, and the objective of these simulations was to determine the impact of $\beta_x$, $\beta_e$, and $\gamma$ on endogenous peak value and peak tick (see Table \ref{t0}). To achieve this, we took a sample of 27000 simulations and analyzed them as described in Algorithm~\ref{a2}.

\begin{algorithm}[htbp]
	
	\begin{algorithmic}[1]
		
		\State Initialize $\beta_x$, $\beta_e$ and $\gamma$ with 30 random values between 0 and 1. 
		\State Initialize the number of susceptible, infected(endogenous and exogenous) and recovered nodes experimentally as: N = 1000000.0, $S_0$ = 999996.0, $I_{x_0}$ = 3.0, $I_{e_0}$ = 1.0 and $R_0$ = 0.0.
		\State For each of the 27000 combinations of $\beta_x$, $\beta_e$, and $\gamma$, with the above initial condition, predict the endogenous and exogenous peak value and peak tick using the Exo-SIR model.
		\State Then, compute the natural logarithm of the peak value and scaled it between 0 and 1.
		\State Finally, fit an OLS Regression Model with $\beta_x$, $\beta_e$ and $\gamma$ as the independent variables and the natural log of the peak value as the dependent variable and analyse the coefficients statistically.
	\end{algorithmic}
	\caption{Algorithm to perform the simulations and analysis by assuming no contact network in the population. \\\textbf{Note:} If we look at the differential equations, the system is not a linear one, but rather exponential. Therefore, we took natural log of the peak value as the dependent variable. Table~\ref{t0} shows the impact of the above three independent variables on the dependent variable.}\label{a2}
\end{algorithm}

The following inferences can be drawn from the results of Regression Analysis.
The p-value for all the three variables is less than 0.05. This means we would reject the null hypothesis and adopt the alternate hypothesis that the impact of all the three parameters on the peak endogenous infection's peak is statistically significant.

The adjusted R-squared value is maximum(0.70) when all the three parameters are considered while fitting the regression model. This means that we can better explain the variation in the dependent variable when considering all three, i.e., $\beta_e$, $\beta_x$, and $\gamma$. Removing any one of them would decrease the adjusted R-squared value. Also, the confidence interval of each parameter is mentioned in Table~\ref{t0}.

$\beta_x$ impacts endogenous infections as much as $\beta_e$(the contribution of both is almost equal), which is an important observation. This means that exogenous factors also have a considerable impact on the endogenous infection, and ignoring the exogenous factors would not give an accurate estimate of the endogenous infections. 

\section{Analysis using real data}
\label{realData}
In this section, we describe the data and the analysis of the implementation of the SIR model and Exo-SIR model on the Covid-19 and Ebola epidemics. 

\subsection{Covid-19 infection in India}
Covid-19 has caused large and persistent negative effects on the world economy\footnote{\url{shorturl.at/qBZ05}}. India is one of the countries that are worst affected. There were many issues that made the spread of Covid-19 in India complicated. One of them was the migration of people from different parts of the country and abroad. 

Many sub-events in India involved the migration of people. Examples are a celebrity coming to India from the UK and socializing at many places even after being tested positive for Covid-19\footnote{\url{shorturl.at/imBGK}}, laborers working in different states or other countries moving back to their native places \cite{rajan2020covid} and large religious meetings with participation from many national and international locations. 

A major sub-event was the Tablighi Jamaat religious congregation in Delhi from $1^{st}$ March $2020$ to $21^{st}$ March $2020$\footnote{\url{https://en.wikipedia.org/wiki/2020_Tablighi_Jamaat_coronavirus_hotspot_in_Delhi}}. Over $9000$ people from various states of India participated in this event\footnote{ \url{shorturl.at/qryKU}}. Nearly $4300$  cases have been reported that can be traced to the event\footnote{\url{shorturl.at/myFQ2}}. As of $18^{th}$ April $2020$, $30\%$ of the cases in India were due to this event\footnote{\url{shorturl.at/iyVY9}}. The number of people from each state is widely deferred. Hence, the impact of the event was significantly different for different states. However, it is reasonable to state that the mobility of people is a causative phenomenon that changed the dynamics of the spread of the virus. 

We apply the Exo-SIR model on a real dataset regarding the spread of the Covid-19 pandemic in the Indian states of Rajasthan, Tamil Nadu, and Kerala from $14^{th}$ March, $2020$ to $14^{th}$ April, $2020$. Exogenous spread dominates endogenous spread in Tamil Nadu, whereas the contrary is true in the case of Rajasthan. Both the endogenous and exogenous spread in Kerala have roughly the exact prevalence. The trends in the analytical study, results of the simulations, and the analysis of the real dataset are consistent. 

We analyzed the data of three states in India, namely Tamil Nadu, Rajasthan and Kerala. The reason for choosing these states is that $i_e \ll i_x$ in Tamil Nadu, $i_e \gg i_x$ in Rajasthan and $i_e \approx i_x$ in Kerala.

We constructed our dataset from three different sources\footnote{The code and all the data used in our experiments will be made openly available upon the acceptance of this paper} for our analysis -- \url{covid19india.org}\footnote{\url{www.covid19india.org}}, the government website of the respective states for their press release to find the daily number of Tablighi cases and Wikipedia page on state-wise daily data\footnote{\url{https://en.wikipedia.org/wiki/Statistics_of_the_COVID-19_pandemic_in_India}}. 

\url{covid19india.org} is a publicly available volunteer-driven dataset of Covid-19 statistics in India\footnote{\url{www.covid19india.org}}. There are multiple files in this dataset. One of which is called \emph{raw data} that captures the anonymized details of the patients. In the raw data, the columns of interest for our study are DateAnnounced, DetectedState, and TypeOfTransmission.

Another file from \url{covid19india.org} is called \emph{states\_daily}. In this file the columns of interest are \emph{states\_daily/status}, \emph{states\_daily/kl}, \emph{states\_daily/rj}, \emph{states\_daily/tn} and \emph{states\_daily/date}. \emph{kl, rj} and \emph{tn} are the codes used in this dataset for the states of Kerala, Rajasthan and Tamil Nadu respectively.

Here, status can have the following values: infected, recovered, and diseased. From these columns, we prepared the time series dataset for each state. The columns available in the dataset we created are daily confirmed, daily deceased, daily recovered, date, total confirmed, total deceased, totally recovered, and daily imported cases.

Another dataset that we used is the compilation of the press releases (news bulletins) from the states' governments under study. This is to get the daily number of cases due to a significant event that influenced the Covid-19 spread in India -- Tablighi Jamaat religious congregation. Since there was no ready made data available, we manually went through the press releases and collected the data.

Now, we discuss how the values in the dataset is mapped on to the variables in the Exo-SIR model. On a particular day, say day $k$, by rearranging and differentiating Equation~\ref{key3}, we get the following.

\begin{equation}
\frac{ds}{dt} = -(\frac{di}{dt} + \frac{dr}{dt})
\end{equation}

\noindent where $\frac{dr}{dt} $ is the sum of the numbers of the daily recovered and the daily deceased cases on day $k$ and

\begin{equation}
\frac{di}{dt} = \frac{di_e}{dt} + \frac{di_x}{dt}
\end{equation}

\noindent where $\frac{di_e}{dt} $ is the daily confirmed cases on day $k$ and  $\frac{di_x}{dt}$ is the sum of daily imported cases on day $k$ and the daily cases due to Tablighi event on day $k$.

The initial values of s, i and r are found as follows.

\begin{equation}
s = 1 - \frac{\mbox{d(0)}}{N}
\end{equation}

where $d(0)$ is the daily confirmed on day $0$ and N is the total population who are prone to the infection. 

$i$ is the total number of confirmed cases on day $0$ and $r$ is the sum of the total numbers of the deceased and the recovered cases on day $0$.

\begin{algorithm}[htbp]
	
	\begin{algorithmic}[1]
		
		\State For each time slice, calculate the values of $ \frac{di}{dt}$ and $\frac{dr}{dt}$ from the dataset.
		\State Consider $s$ as the susceptible people from the population of the state under study. 
		\State Calculate the cumulative values $i$ and $r$. 
		\State Find $\gamma$, $\beta_e$ and $\beta_x$ using values of the time for which the data is available.
		\State Run the Exo-SIR model with these values as the initial values and plot $i_e$ in the presence of $i_x$ and $i_e$ in the absence of $i_x$
	\end{algorithmic}
	\caption{Algorithm to plot the Exo-SIR model.}\label{a1} 
\end{algorithm} 

Next, we analyze the data from Tamil Nadu, Rajasthan, and Kerala. We compare the peak tick and peak value of the plot of $i_e$ in the presence and absence of $i_x$. This would give information about the impact of $i_x$ on $i_e$. For this purpose, we used Algorithm~\ref{a1}.

\begin{figure}[htbp]
	\centering
	\includegraphics[width=.5\textwidth]{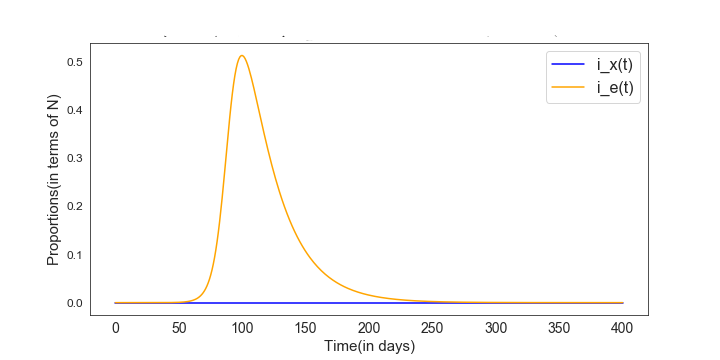}
	\caption{$I_e$ in the presence of $i_x$. The values of $i_x$ are very small for the scale of this plot. Hence it is plotted separately. Please refer the Figure~\ref{f108}}
	\label{f110}
\end{figure}

\begin{figure}[htbp]
	\centering
	\includegraphics[width=.5\textwidth]{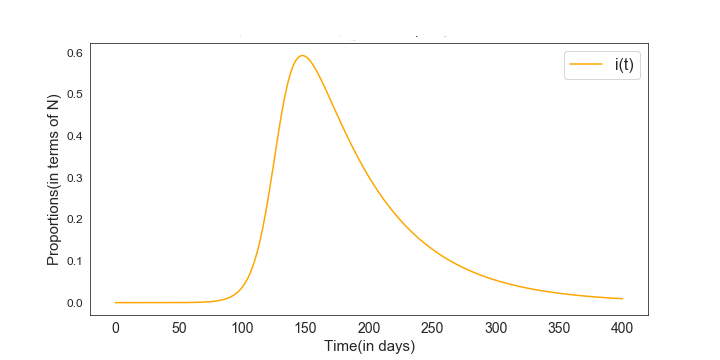}
	\caption{$I_e$ in the absence of $i_x$ }
	\label{f109}
\end{figure}

\begin{figure}[htbp]
	\centering
	\includegraphics[width=.5\textwidth]{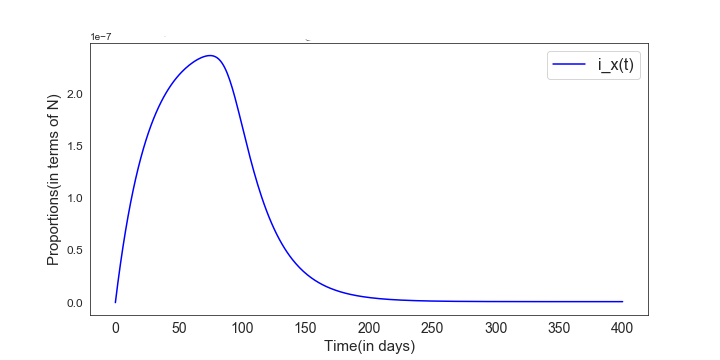}
	\caption{$i_x$ in Exo-SIR model. Please note that y axis is in the scale of $10^{-4}$.}
	\label{f108}
\end{figure}

\begin{figure}[htbp]
	\centering
	\includegraphics[width=.5\textwidth]{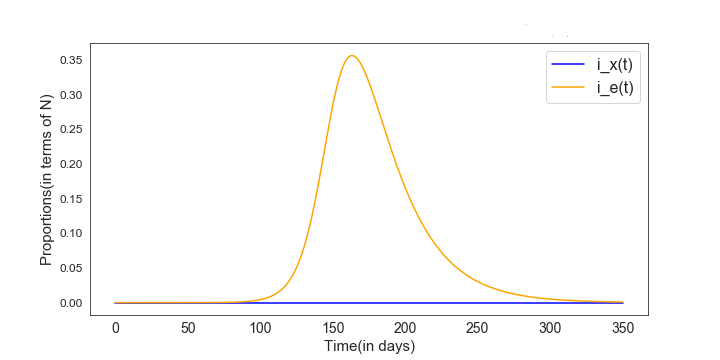}
	\caption{$I_e$ in the presence of $i_x$. The values of $i_x$ are very small for the scale of this plot. Hence it is plotted separately. Please refer the Figure~\ref{f105}}
	\label{f107}
\end{figure}

\begin{figure}[htbp]
	\centering
	\includegraphics[width=.5\textwidth]{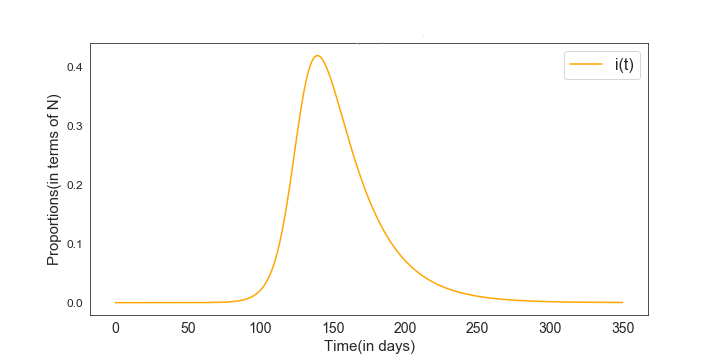}
	\caption{$I_e$ in the absence of $i_x$ }
	\label{f106}
\end{figure}

\begin{figure}[htbp]
	\centering
	\includegraphics[width=.5\textwidth]{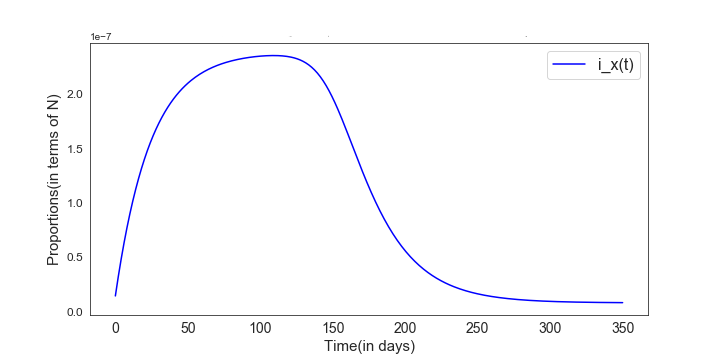}
	\caption{ $i_x$ in Exo-SIR model. Please note that the y axis is in the scale of $10^{-6}$.}
	\label{f105}
\end{figure}

\begin{figure}[htbp]
	\centering
	\includegraphics[width=.5\textwidth]{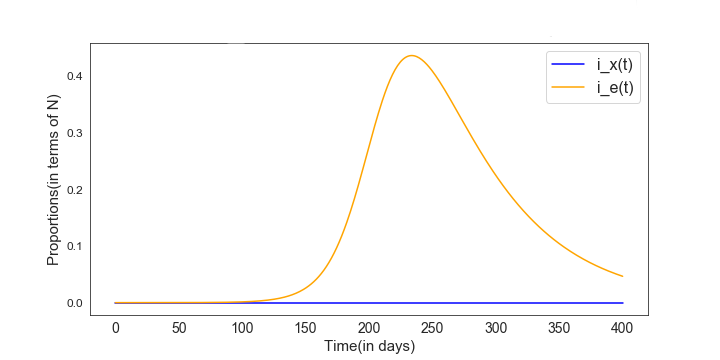}
	\caption{$I_e$ in the presence of $i_x$ in the state of Kerala.  The values of $i_x$ are very small for the scale of this plot. Hence it is plotted separately. Please refer the Figure~\ref{f101}}
	\label{f103}
\end{figure}

\begin{figure}[htbp]
	\centering
	\includegraphics[width=.5\textwidth]{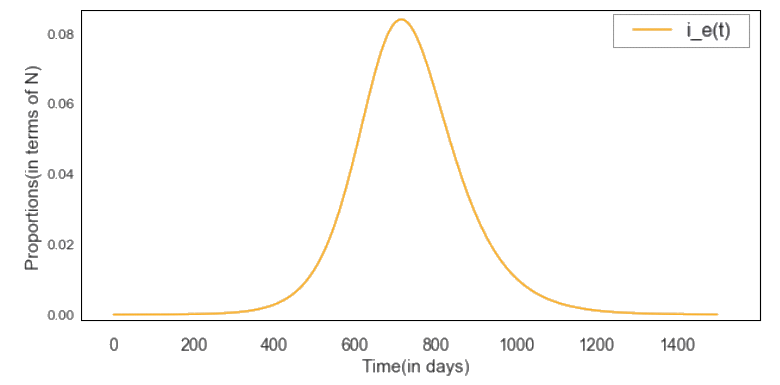}
	\caption{$I_e$ in the absence of $i_x$ in the state of Kerala }
	\label{f102}
\end{figure}

\begin{figure}[htbp]
	\centering
	\includegraphics[width=.5\textwidth]{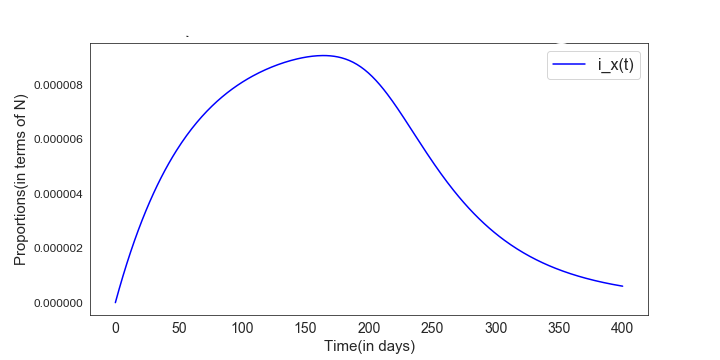}
	\caption{$i_x$ in Exo-SIR model. Please note that the y axis is in the scale of $10^{-5}$.}
	\label{f101}
\end{figure}

For the state of Tamil Nadu, the plots of $I_e$ in the presence and absence of $i_x$ are plotted in Figure~\ref{f110} and Figure~\ref{f109} respectively. For the state of Rajasthan, the plots of $I_e$ in the presence and absence of $i_x$ are plotted in Figure~\ref{f107} and Figure~\ref{f106} respectively. For the state of Kerala, the plots of $I_e$ in the presence and absence of $i_x$ are plotted in Figure~\ref{f103} and Figure~\ref{f102}, respectively. In all these plots, we can see that $i_x$ is very small compared to $i_e$. Yet, $i_x$ is having an impact on $i_e$. $I_x$ is plotted separately in Figure~\ref{f108}, Figure~\ref{f105} and Figure~\ref{f101}.

\begin{table}[htbp]
	\caption{Impact of $I_x$ on $I_e$ in the state of Tamil Nadu}
	\label{t1}
	\begin{tabular}{|p{.15\textwidth}|l|l|}
		\hline
		&&\\
		& peak value & peak tick \\
		&&\\
		\hline
		&&\\
		$I_e$ in the presence of $i_x$ & $0.1714$ & $907$ Days \\
		&&\\
		$I_e$ in the absence of $i_x$ & $0.1710$ & $1351$ Days\\
		&&\\
		\hline
	\end{tabular}
\end{table}

\begin{table}[htbp]
	\caption{Impact of $I_x$ on $I_e$ in the state of Rajasthan}
	\label{t2}
	\begin{tabular}{|p{.15\textwidth}|l|l|}
		\hline
		&&\\
		& peak value & peak tick \\
		&&\\
		\hline
		&&\\
		$I_e$ in the presence of $i_x$ & $0.3487077$ & $143$ Days\\
		&&\\
		$I_e$ in the absence of $i_x$ & $0.3486663$ & $147$ Days\\
		&&\\
		\hline
	\end{tabular}
\end{table}

\begin{table}[htbp]
	\caption{Impact of $I_x$ on $I_e$ in the state of Kerala}
	\label{t3}
	\begin{tabular}{|p{.15\textwidth}|l|l|}
		\hline
		&&\\
		& peak value & peak tick \\
		&&\\
		\hline
		&&\\
		$I_e$ in the presence of $i_x$ & $0.0842$ & $608$ Days\\
		&&\\
		$I_e$ of in the absence of $i_x$ & $0.0841$ & $715$ Days\\
		&&\\
		\hline
	\end{tabular}
\end{table}

The peak tick and peak values corresponding to the $I_e$ of the Exo-SIR model in the presence and absence of $i_x$ for Tamil Nadu, Rajasthan, and Kerala are mentioned in Table~\ref{t1}, Table~\ref{t2} and Table~\ref{t3} respectively. In all the tables, we can see that the peak value of $i_e$ is different when the case of $i_x$ is present. Also, we can see that the peak tick of $i_e$ is different for the instance when $i_x$ is present.

\begin{figure}[htbp]
	\centering
	\includegraphics[width=.5\textwidth]{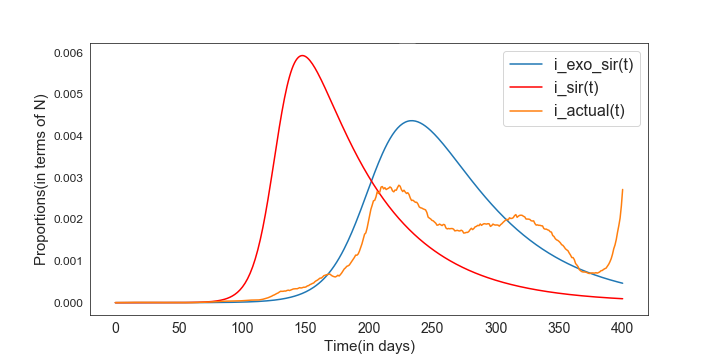}
	\caption{Comparison of the predictions of Exo-SIR and SIR models with real data for Covid-19 in Kerala}
	\label{pic6.1.1}
\end{figure}

\begin{figure}[htbp]
	\centering
	\includegraphics[width=.5\textwidth]{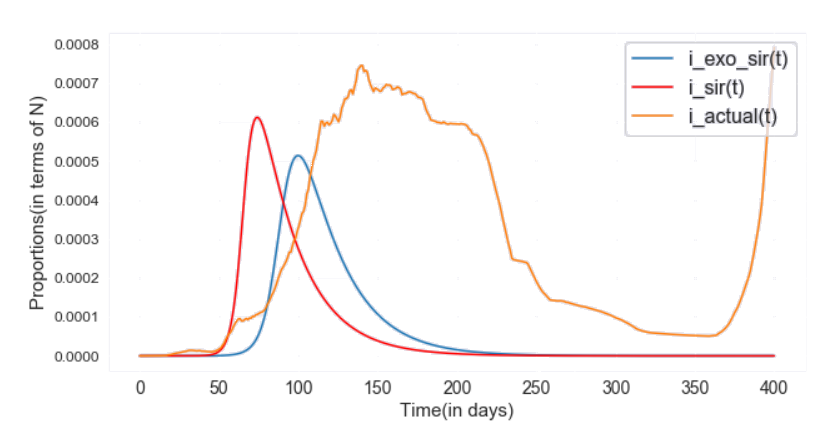}
	\caption{Comparison of the predictions of Exo-SIR and SIR models with real data for Covid-19 in Tamil Nadu}
	\label{pic6.1.2}
\end{figure}

\begin{figure}[htbp]
	\centering
	\includegraphics[width=.5\textwidth]{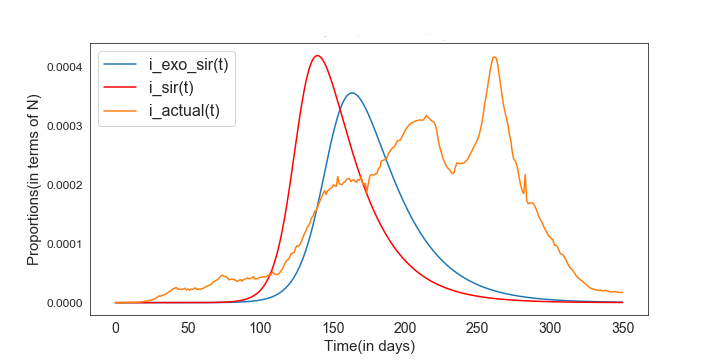}
	\caption{Comparison of the predictions of Exo-SIR and SIR models with real data for Covid-19 in Rajasthan}
	\label{pic6.1.3}
\end{figure}

Finally, we present the comparison of the predictions of Exo-SIR model and SIR model with the real data for the following cases:
\begin{enumerate}
    \item Covid-19 in Kerala (Figure \ref{pic6.1.1})
    \item Covid-19 in Tamil Nadu (Figure \ref{pic6.1.2})
    \item Covid-19 in Rajasthan (Figure \ref{pic6.1.3})
\end{enumerate}

Here, the peak values are scaled down as they are very high for both SIR and Exo-SIR predictions. This may be due to the fact that in both SIR and Exo-SIR models, we assume that each infected person is equally likely to infect all the susceptible people. In the real life, this is not true. However, we can see that in all the three cases (shown in Figure \ref{pic6.1.1}, \ref{pic6.1.2} and \ref{pic6.1.3}), the peak of the Exo-SIR model is closer to the peak of the real data.

\subsection{Covid-19 infection in the USA}

In this section, we discuss the analysis that we carried out on the data of Covid-19 infection in the USA.

We constructed our dataset from two different sources\footnote{The code and all the data used in our experiments will be made openly available upon the acceptance of this paper} for our analysis -- \url{kaggle.com} and incoming tourists travel data for the USA from the CEIC database \footnote{\url{https://www.ceicdata.com/en/indicator/united-states/visitor-arrivals}}. 

Now, we discuss how the values in the dataset is mapped on to the variables in the Exo-SIR model. We calculated the number of endogenous infections $(I_E(t))$ from the following equation.
\begin{equation}
I_E(t)\ =\ I_E(t-1) + Daily(t)-D(t-1)
\end{equation}

where, $Daily(t)$ is the daily new cases at the time slice $t$ and $D(t-1)$ is the deaths from within the USA population at the time slice $t-1$.

We estimated infected tourists death number from endogenous deaths in the following way. First, we calculated $\gamma$ from endogenous data by using equation

\begin{equation}
\gamma = \frac{dr/dt}{i}
\end{equation}

Applied the same gamma to get the number of deaths from data of exogenous infections using the equation
\begin{equation}
r(t) = r(t-1) + dr/dt
\end{equation}

where 
\begin{equation}
dr/dt = \gamma*i(t-1)
\end{equation}

Then we calculated the number of exogenous infections $(I_X(t))$ by using the equation:

\begin{equation}
I_X(t) = I_X(t-1)\ + Daily(t) - D(t)
\end{equation}

where $Daily(t)$ is the daily new tourist cases at the time slice $t$ and $D(t)$ is the number of deaths at the time slice $t$

Then we calculated the number of susceptible people by using the following equation:

\begin{equation}
S(t) = N-I_E^c(t)-I_X^c(t)
\end{equation}

where $I_E^c(t)$ is the cumulative value of $I_E(t)$ and $I_X^c(t)$ is the cumulative value of $I_X(t)$.

Finally we computed $ \frac{d(i_e)}{dt}$, $\frac{d(i_x)}{dt}$, $\frac{d(r)}{dt}$ and $\frac{d(s)}{dt}$ values.

\begin{figure}[htbp]
	\centering
	\includegraphics[width=.5\textwidth]{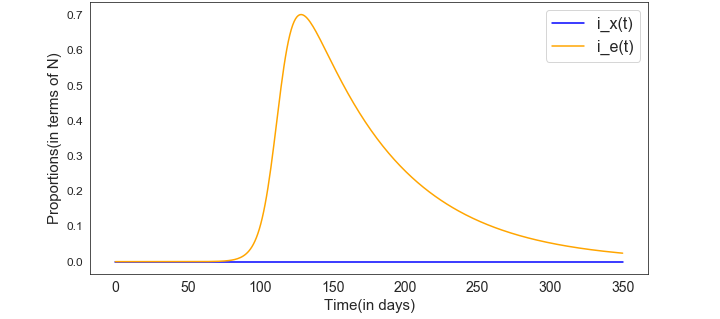}
	\caption{$i_e$ and $i_x$ for Covid-19 in the USA}
	\label{pic11USA}
\end{figure}

\begin{figure}[htbp]
	\centering
	\includegraphics[width=.5\textwidth]{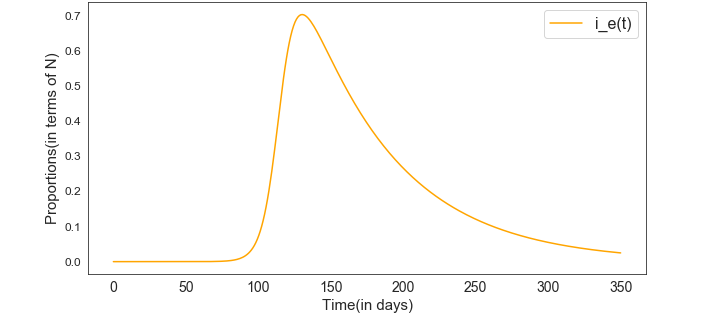}
	\caption{$i_e$ in the absence of $i_x$ for Covid-19 in the USA}
	\label{pic13USA}
\end{figure}

\begin{figure}[htbp]
	\centering
	\includegraphics[width=.5\textwidth]{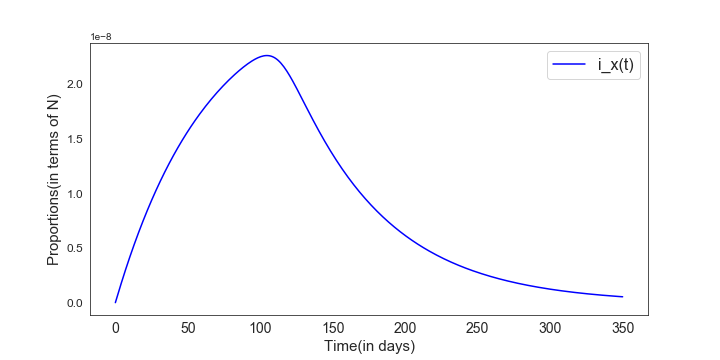}
	\caption{$i_x$ for Covid-19 in the USA}
	\label{pic14USA}
\end{figure}

Next, we analyze the Covid-19 data from the USA by applying the Exo-SIR model. We compare the peak tick and peak value of the plot of $i_e$ in the presence and absence of $i_x$. This would give information about the impact of $i_x$ on $i_e$. For this purpose, we used Algorithm~\ref{a1}. The cases in the presence and absence of $I_x$ is plotted in Figure~\ref{pic11USA} and \ref{pic13USA} respectively. $I_x$ is plotted in the Figure~\ref{pic14USA}.

In these plots, it can be observed that the peak and the height of the peak are different compared to the values in the absence of $i_x$ The peak tick and peak values corresponding to the $I_e$ of Exo-SIR model in the presence and absence of $i_x$ are mentioned in Table~\ref{t3USA}.

\begin{table}[htbp]
	\caption{Impact of $I_x$ on $I_e$ in Covid-19 in the USA}
	\label{t3USA}
	\begin{tabular}{|p{.15\textwidth}|l|l|}
		\hline
		&&\\
		& peak value & peak tick \\
		&&\\
		\hline
		&&\\
		$I_e$ in the presence of $i_x$ & 0.7 &  130 days \\
		&&\\
		$I_e$ in the absence of $i_x$ & 0.7 & 135 days \\
		&&\\
		\hline
	\end{tabular}
\end{table}

\begin{figure}[htbp]
	\centering
	\includegraphics[width=.5\textwidth]{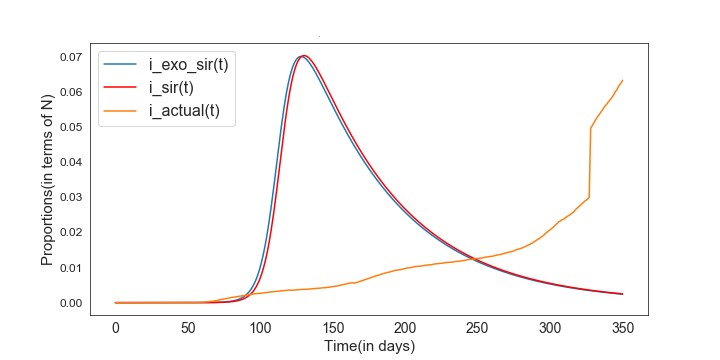}
	\caption{Comparison of the predictions of Exo-SIR and SIR models with real data for Covid-19 in the USA}
	\label{pic6.1.USA}
\end{figure}

Figure~\ref{pic6.1.USA} shows the comparison of the predictions of Exo-SIR model and SIR model with the real data. Here, we can see that the peaks in the SIR and Exo-SIR plots are of the same height and are coming more or less simultaneously. However, both of them are very different from the peak position in the real data.

\subsection{Ebola infection in Guinea}

Ebola, also known as EVD, was another severe, often fatal epidemic that hit the Western African countries from 2014 to 2016, particularly Guinea, Sierra Leone, and Liberia. Its fatality rate\footnote{\url{https://www.who.int/health-topics/ebola}} varies from 25\% to 90\%. Like the case of Covid-19, there was migration of people from abroad, especially tourists traveling into these countries. The dataset regarding travel and tourism is publicly available\footnote{\url{https://www.unwto.org/unwto-tourism-dashboard}}.

We compared peak tick and peak value of the plot of $i_e$ in the presence and absence of $i_x$, as per Algorithm~\ref{a1}. This gave us information and important insights on the impact of $i_x$ on $i_e$.

We constructed our dataset from two different sources: \url{kaggle.com} and incoming tourists travel data for Guinea from UNWTO Dashboard\footnote{\url{https://www.unwto.org/seasonality}}. 

Now, we discuss how the values in the dataset is mapped on to the variables in the Exo-SIR model. We calculated the number of endogenous infections $(I_E(t))$ from the following equation.

\begin{equation}
I_E(t)\ =\ I_E(t-1) + M(t)-D(t-1)
\end{equation}

where $M(t)$ is the monthly new cases at the time slice $t$ and $D(t-1)$ is the deaths from within the Guinea population at the time slice $t-1$.

We estimated infected tourists death number from endogenous deaths in the following way. First, we calculated $\gamma$ from endogenous data by using the equation 

\begin{equation}
\gamma = \frac{dr/dt}{i}
\end{equation}

Then we applied the same gamma to get the number of deaths from data of exogenous infections using the equation:

\begin{equation}
r(t) = r(t-1) + dr/dt
\end{equation}

where 
\begin{equation}
dr/dt = \gamma*i(t-1)
\end{equation}

Then we calculated number of exogenous infections $(I_X(t))$ by using the equation:

\begin{equation}
I_X(t) = I_X(t-1)\ + M(t) - D(t)
\end{equation}

where $M(t)$ is the monthly new tourist cases at the time slice $t$ and $D(t)$ is the number of deaths at the time slice $t$.

Then we calculated the number of susceptible people by using the following equation:

\begin{equation}
S(t) = N-I_E^c(t)-I_X^c(t)
\end{equation}

where $I_E^c(t)$ is the cumulative value of $I_E(t)$ and $I_X^c(t)$ is the cumulative value of $I_X(t)$.

Finally we computed $ \frac{d(i_e)}{dt}$, $\frac{d(i_x)}{dt}$, $\frac{d(r)}{dt}$ and $\frac{d(s)}{dt}$ values.

\begin{figure}[htbp]
	\centering
	\includegraphics[width=.5\textwidth]{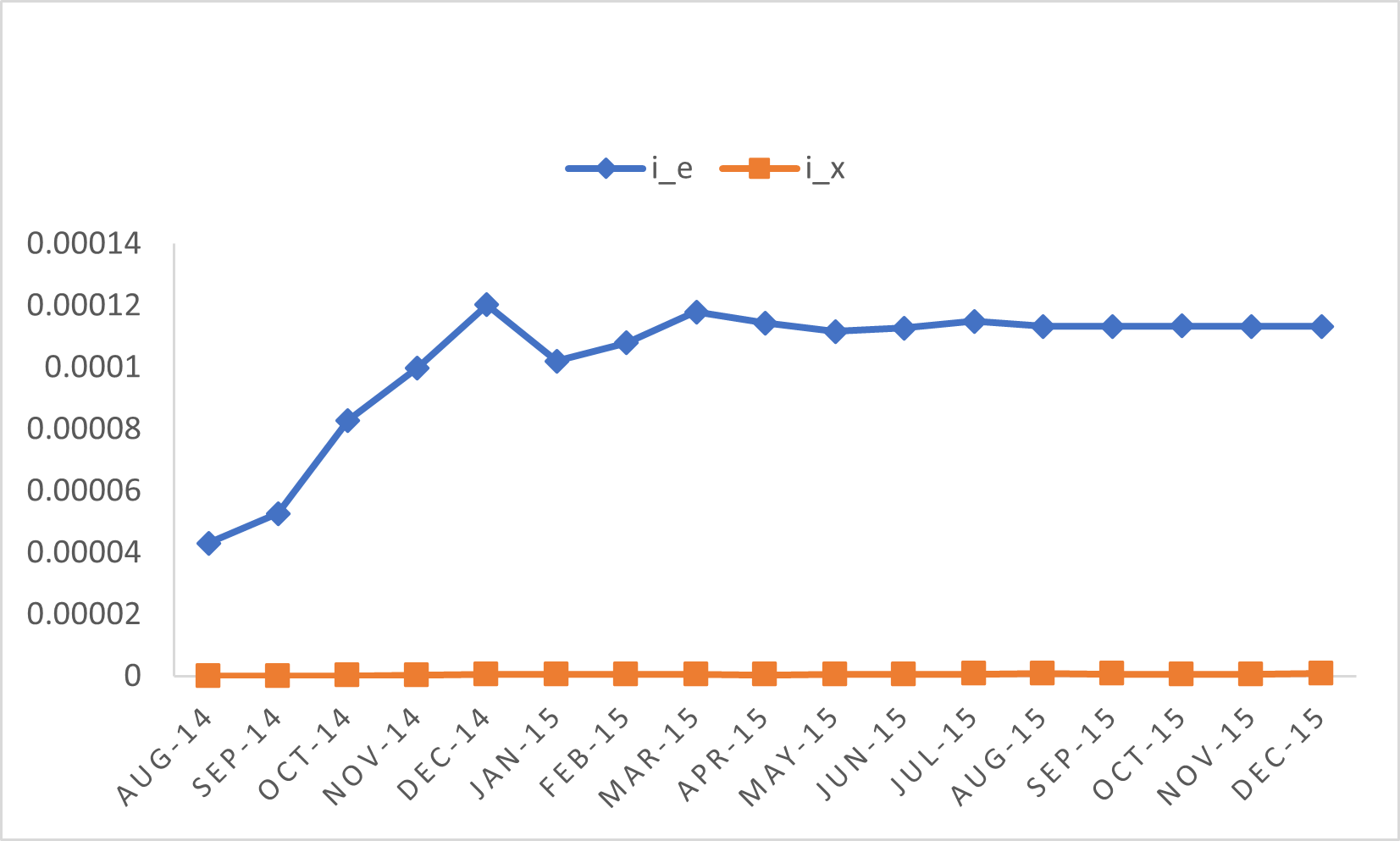}
	\caption{$i_e$ and $i_x$ for Ebola}
	\label{pic11}
\end{figure}

\begin{figure}[htbp]
	\centering
	\includegraphics[width=.5\textwidth]{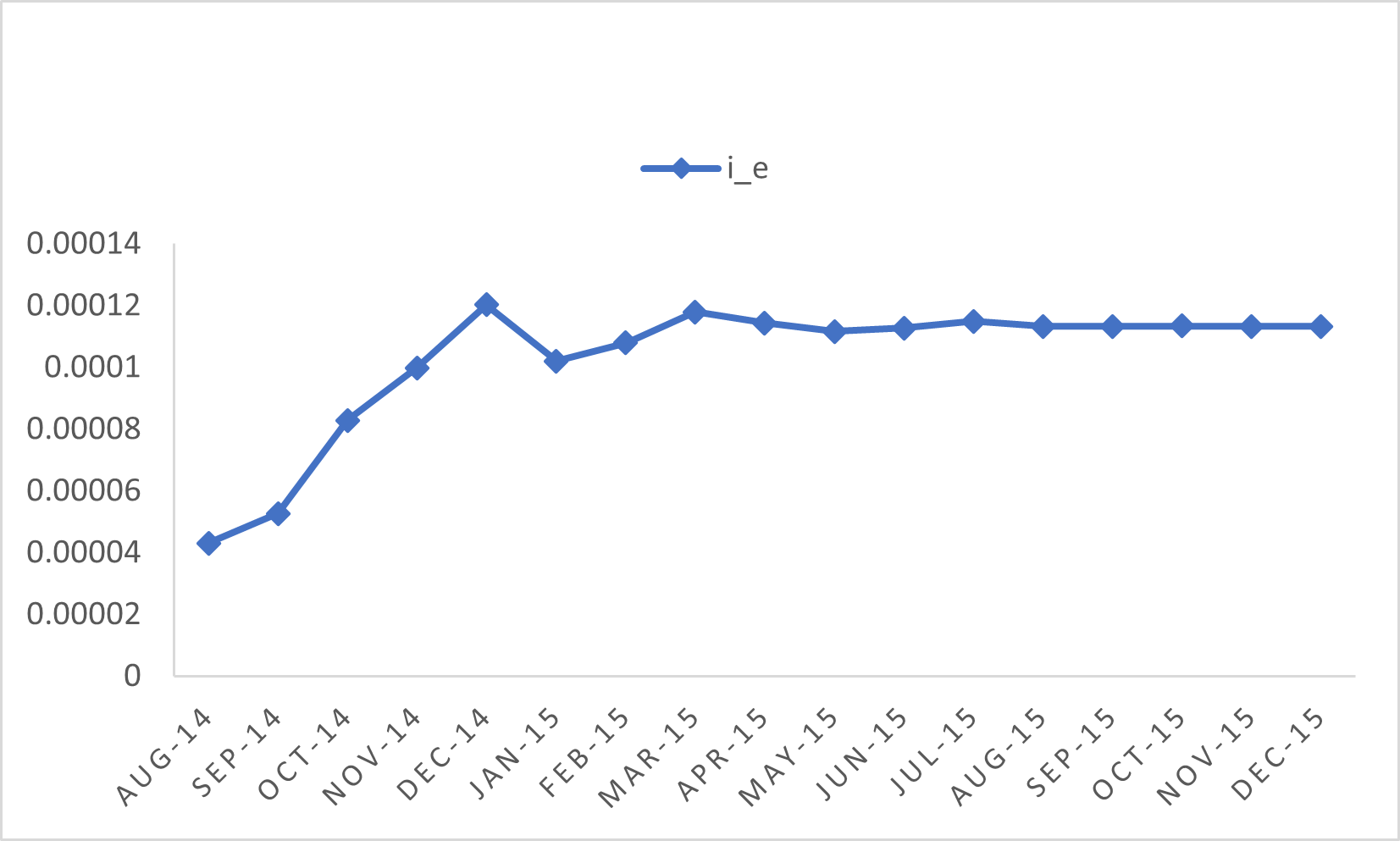}
	\caption{$i_e$ in the absence of $i_x$ for Ebola}
	\label{pic13}
\end{figure}

\begin{figure}[htbp]
	\centering
	\includegraphics[width=.5\textwidth]{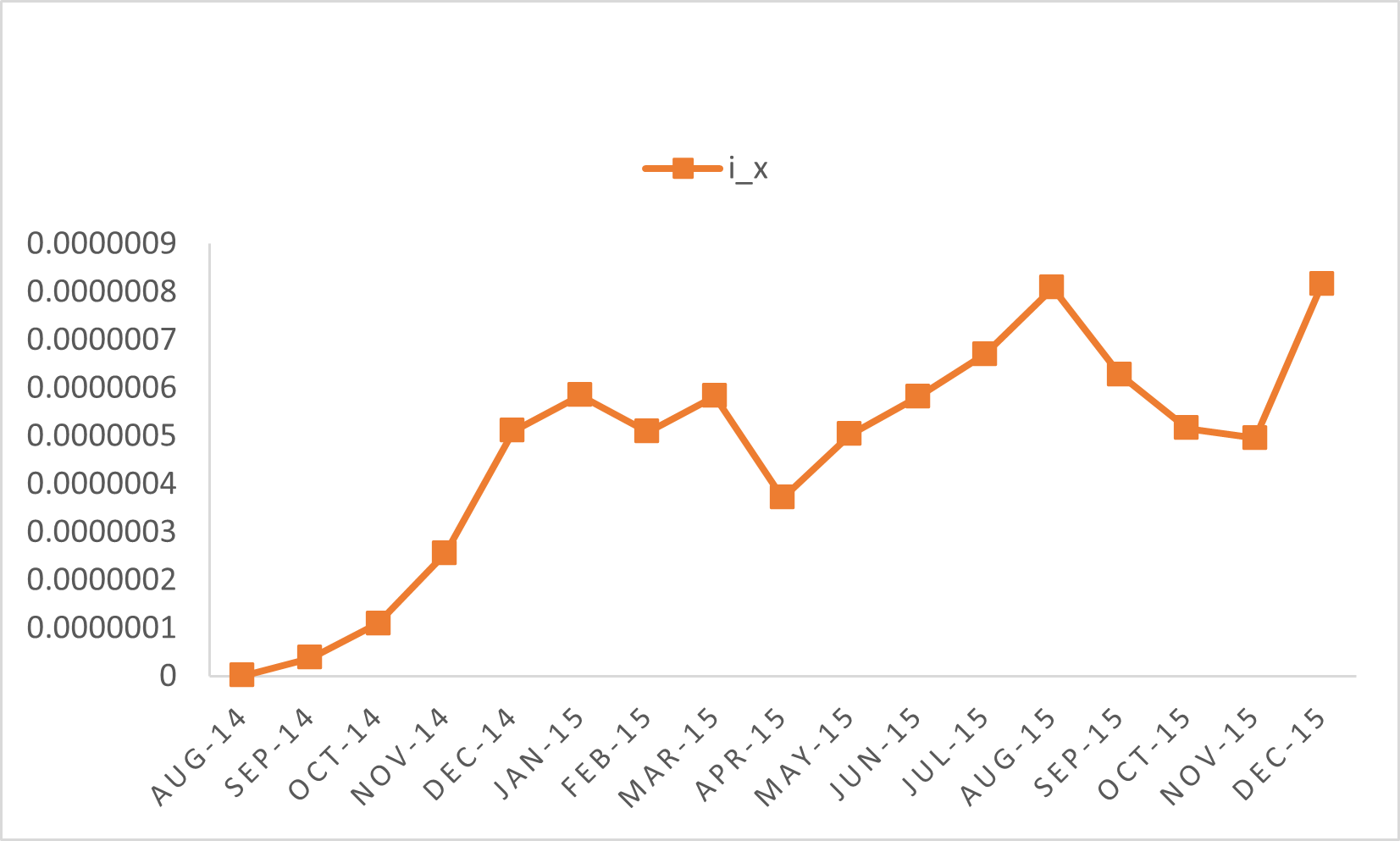}
	\caption{$i_x$ for Ebola}
	\label{pic14}
\end{figure}

Next, we analyze the data from Guinea. We compare the peak tick and peak value of the plot of $i_e$ in the presence and absence of $i_x$. This would give information about the impact of $i_x$ on $i_e$. For this purpose, we used Algorithm~\ref{a1}. The cases in the presence and absence of $I_x$ are plotted in Figure~\ref{pic11} and \ref{pic13} respectively. $I_x$ is shown in Figure~\ref{pic14}.

From Figure ~\ref{pic11}, \ref{pic13}, and \ref{pic14}, the peak and the height of the peak are different compared to the values in the absence of $i_x$. The peak tick and values corresponding to $I_e$ of the Exo-SIR in the presence and absence of $i_x$ are mentioned in Table~\ref{t3Ebola}.

\begin{table}[htbp]
	\caption{Impact of $I_x$ on $I_e$ in Guinea}
	\label{t3Ebola}
	\begin{tabular}{|p{.15\textwidth}|l|l|}
		\hline
		&&\\
		& peak value & peak tick \\
		&&\\
		\hline
		&&\\
		$I_e$ in the presence of $i_x$ & 0.00012 &  5 Months\\
		&&\\
		$I_e$ in the absence of $i_x$ & 0.00012 & 6 Months\\
		&&\\
		\hline
	\end{tabular}
\end{table}

\begin{figure}[htbp]
	\centering
	\includegraphics[width=.5\textwidth]{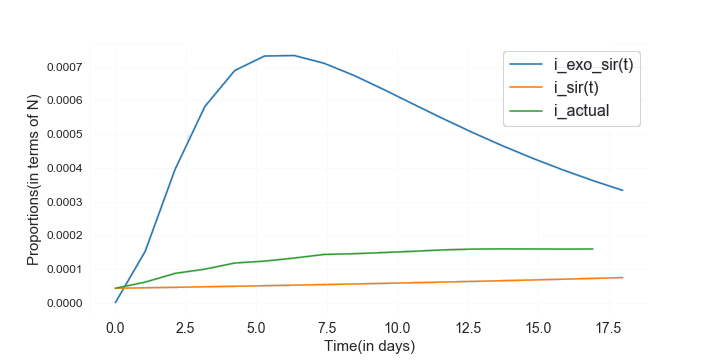}
	\caption{Comparison of the predictions of Exo-SIR and SIR models with real data for Ebola in Guinea}
	\label{pic6.1.Ebola}
\end{figure}

Figure~\ref{pic6.1.Ebola} shows the comparison of the predictions of Exo-SIR model and SIR model with the real data. Here, we can see that the peak of Exo-SIR and SIR models are coming differently and they are coming far from the peak of the actual data.

\subsection{Discussion}

Both Covid-19 and Ebola satisfy our hypothesis that the endogenous spread changes in the presence of exogenous spread. Also, the results in the case of Covid-19 infection in India show that the Exo-SIR model predicts the epidemic's peak tick better than the SIR model. 

Covid-19 in the USA and Ebola in Guinea show less accurate predictions than Covid-19 in India. This may be because of the following reasons.

In these cases, we took the data from the beginning of the spread of the infection. As soon as the infections started growing, the governments began multiple interventions to curb the spread of the epidemics. If these efforts were successful, that would change the values of the constants that we calculated using the initial values. This will reflect in the curve of the real data primarily by delaying the peak and flattening the curve. This can be observed in the real data of Covid-19 in the USA and Ebola in Guinea. On the other hand, in the case of the data from India, we took the data when the migration of people after the Tablighi religious congregation happened. By this time, India was already on the alert, and the government had already intervened in the matter. Hence, our calculation of the constants was closer to the actual values.

We analyzed a sub-event in the case of Covid-19 in India, the Tablighi religious congregation, with many participants from almost all the states in India. The number of these people who traveled back to the states was considered $I_x$. The probability of these people being infected was very high as the event was a hot spot of the infection. However, in the case of Covid-19 in the USA and Ebola in Guinea, we considered the tourist arrival data as $I_x$. We made strong assumptions in these cases due to the unavailability of the daily inflow of the infected people to the population. In the case of Covid-19 in the USA, we calculated the external infection as the tourist arrival data multiplied by the total infection in the world. We normalized it by the total population of the world. In the case of Ebola in Guinea, we calculated the external infection as the tourist arrival data multiplied with the total population of the three countries where the infection was the most prevalent and normalized it by the world's total population. In these cases, the probability that all the people in the travel data are infected is comparatively less. This may be the reason for the difference. It is important to note that the SIR model performed equally bad in these cases. This also suggests that the issue might be with the data.

The peak value of the predictions of both SIR and Exo-SIR models was very high compared to the real values. The reason for this may be the following. In the case of SIR and Exo-SIR models, we assume that susceptible people are equally likely to get infected from each infected person in the population. This is not true in real life. In real life, people are likely to get infected only from those they contact. This number is much less than the assumption in both SIR and Exo-SIR models.

\section{Conclusion}

This study introduced the Exo-SIR model by extending the SIR model. Unlike the other epidemiological models, the Exo-SIR model differentiates between the endogenous and exogenous spread of virus/information. We studied the model in the following ways:

\begin{enumerate}
	\item Analytical study 
	\item Simulation considering the presence of contact network of the population ans assuming it to be a scale free network
	\item Simulation without considering the presence of contact network
	\item Implementation of the Exo-SIR model on real data about the spread of Covid-19 in India, Covid-19 in the USA and the spread of Ebola in Guinea.
\end{enumerate}

We found that all the four analyses mentioned here converge to the same result: the peak comes differently in time and size when the exogenous source is present. We studied the impact of exogenous infection on endogenous diffusion. We found that exogenous diffusion impacts the endogenous spread of infection. If there are exogenous sources of infection, like in the case of Covid-19 or Ebola, then the Exo-SIR model is more appropriate to estimate the scenario better. This will help the government allocate its resources better as the endogenous and exogenous spread needs different sets of actions to stop them.

\noindent\textbf{Limitations and Future works:} We used the SIR model for comparison as it is simple and widely used. Other models like SEIR, SEYAR, etc. that could be used for a similar study. There is scope for introducing the external source of infection to these models like SEIR and SEYAR. Also, we have considered only one external source of infection. There may exist multiple external sources of infection like bats, pigs, birds, etc. Another possible scenario is the possible presence of multiple viruses. We propose to study these in the future. 

\section{Acknowledgement}

Amit Sheth and Manas Gaur are supported by the National Science Foundation (NSF) Award 2133842, "EAGER: Advancing Neuro-symbolic AI with Deep Knowledge-infused Learning." Any opinions, findings, and conclusions/recommendations expressed in this material are those of the author(s)  and do not necessarily reflect the views of the NSF.

We thank Melissa Nolan and Stella Self from Arnold School of Public Health, the University of South Carolina, for their valuable comments and helpful suggestions. In addition, we thank  Ch V Radhasai Rupesh for helping in data collection.

\bibliography{sn-article}

\end{document}